# Large-field-of-view Chip-scale Talbot-grid-based Fluorescence Microscopy

Shuo Pang[a,1,2], Chao Han[a,1], Mihoko Kato[c], Paul W. Sternberg[c], Changhuei Yang[a,b].

[a]Department of Electrical Engineering, California Institute of Technology, Pasadena, CA, 91125, USA

[b]Department of Bioengineering, California Institute of Technology

[c]HHMI and Division of Biology, California Institute of Technology

[1]S.P. and C.H. contribute to this work equally.

[2]To whom correspondence should be addressed. E-mail: spang@caltech.edu

Abstract:

The fluorescence microscope is one of the most important tools in modern clinical diagnosis and biological science. However, its expense, size and limited field-of-view (FOV) are becoming bottlenecks in key applications such as large-scale phenotyping and low-resource-setting diagnostics. Here we report a low-cost, compact chip-scale fluorescence-imaging platform, termed the Fluorescence Talbot Microscopy (FTM), which utilizes the Talbot self-imaging effect to enable efficient fluorescence imaging over a large and directly-scalable FOV. The FTM prototype has a resolution of 1.2 μm and an FOV of 3.9×3.5 mm$^2$. We demonstrate the imaging capability of FTM on fluorescently labeled breast cancer cells (SK-BR-3) and HEK cells expressing green fluorescent protein.

Introduction

Due to the high specificity and high sensitivity of fluorescent probes, fluorescence microscopy plays a vital role in modern clinical diagnosis and biological research. To name a few examples: high specificity fluorescent probes against cancer markers are used for cancer diagnosis applications (1); encoded fluorescent proteins enable minimally invasive observation of gene expression for studying functional genomics and stem cell differentiation (2, 3) ; specialized fluorophores can also serve as indicators to visualize excitable cell firing and calcium transients for functional studies in neurons (4).

As the dominant instrument in these applications, the conventional fluorescence microscope is becoming a bottleneck in these rapidly emerging and evolving areas. A conventional fluorescence microscope is bulky, expensive and has a limited field-of-view (FOV). In recognition of these disadvantages, several technologies have recently been developed to enable high throughput screening for diagnosis and gene activity quantification (5-7) . These systems work by optimizing the transport of samples under the conventional microscope's FOV to increase the throughput. However, the inclusion of a conventional microscope precludes system parallelization that can potentially bring orders-of-magnitude improvement in throughput.

The FOV of the conventional microscope objective is inversely related to the resolution. As a point of reference, a standard 20X microscope objective (Plan, Olympus) has a resolution ~ 0.8 μm, a FOV of ~ $5 \times 10^{-2}$ mm$^2$ and a numerical aperture (N.A.) of 0.4. The N.A. is a measure of the objective's light collection solid angle. The objective's resolution is also constrained by the its N.A.. The link between the FOV and the resolution is an inherent disadvantage of the conventional microscope, and is primarily due to the aberrations associated with optical elements in the microscope's objective.

Over the past few years, there have been a number of significant efforts in rethinking the implementation of fluorescence microscopy to circumvent this limitation (8-10) . A lens-less microscopy method based on post digital processing is a good illustrative example (8, 9). This method lays the sample above a sensor chip with an absorptive filter in between. The samples are excited with uniform illumination. The relatively large distance between the fluorophore and the sensor results in a significant fluorescence spread on the sensor (~50 μm) (9). A distribution map of the fluorophores based on the raw image is computed. However, this method requires precise measurements of fluorescence point spread function (PSF) (11). Furthermore, the requirement on fluorophore distribution sparsity prevents this method from being useful for general fluorescence imaging applications (12). For example, in the case of a standard microscope slide sample, the effective resolution of this approach cannot exceed the PSF, which is of the orders of 10's of microns.

A different chip-scale approach, the fluorescence optofluidic microscope (FOFM) method (10), resolves this difficulty by employing Fresnel Zone Plates to form an array of excitation focal spots. As a sample flows through the slanted channel and across the focal array, each spot effectively performs a line scan across the sample. By spacing the light spots well apart from each other, the fluorescence signal from each spot can be measured without cross talk, even if the filter layer is thick. The image resolution is ultimately determined by the focal spot size (~ 0.6 μm full width at half maximum). The main disadvantage of this method is that it can work only with samples that can be suspended in a fluid medium.

In this work, we demonstrate a fluorescence chip-scale microscopy method, termed Fluorescence Talbot Microscopy (FTM), which is based on the Talbot effect. This method can achieve microscopic resolution, and can image both suspended and adherent cells.

The FTM method utilizes the Talbot effect to project a grid of focused excitation light spots onto the sample. The sample rests on a filter-coated CMOS sensor chip. The fluorescence emissions associated with each focal spot are collected by the sensor chip and are composed into a sparsely sampled fluorescence image. By raster scanning the Talbot focal spot grid across the sample and collecting a sequence of sparse images, we can then reconstruct a filled-in high-resolution fluorescence image.

Our FTM prototype has a demonstrated resolution of 1.2 μm, an FOV of 3.9 × 3.5 mm$^2$, and a collection efficiency comparable to a ~ 0.7 N.A. conventional microscope objective.

Here we first present our prototype design and explain the imaging principle in detail. Next, we establish the resolution of the FTM prototype. Finally, we demonstrate the full FOV imaging capability of the FTM on human breast cancer cells (SK-BR-3) labeled with quantum dots and live HEK cells expressing green fluorescent protein (GFP).

## Results

**System design of FTM**

FTM takes advantage of the recent development and commercialization effort of semiconductor and Micro-electro-mechanical System (MEMS). The FTM prototype utilizes a microlens grid (SUSS, 18-00407) to transform the incident coherent plane wave with wavelength $\lambda$ into a grid of tightly focused light spots at the lenses' focal distance $Z_f$=90 µm. The separation between adjacent focal spots is equal to the microlens grid pitch $d$=30 µm. This light grid then propagates a distance of $z=nZ_T$, where n is an integer and $Z_T = 2d^2/\lambda$ is defined as the Talbot length, and reforms into focused light spots. This self-imaging phenomenon is known as the Talbot effect (13). For the FTM prototype, we employ a laser (Fibertech, $\lambda$=488 nm, 30 mW) to serve as the light source. This allows us to create a focused light spot grid (at the first Talbot image, $n$=1) at a distance of 3.7 mm below the microlens grid.

The Talbot effect homogenizes the intensity of the periodic input field. In a prior work, we have applied the Talbot effect in bright field microscopy and demonstrated that this Talbot focal grid has good focusing characteristics (~1 µm spot diameter) and spot power uniformity (14).

In the FTM method, we project the grid of Talbot focused spots onto the target sample. The sample rests on a filter-coated CMOS sensor chip (Fig. 1(a)). The fluorescence emissions associated with each focal spot are collected by the sensor chip and are composed into a sparsely sampled fluorescence image. By angularly tilting of the incident laser (Fig. 1(b)), we can then raster scan the Talbot focal spot grid across the sample and collect a sequence of sparse images. Finally, we reconstruct a filled-in high-resolution fluorescence image (Fig. 1(c)).

At first glance, the use of the Talbot phenomenon may seem superfluous as one can use the original focused spots from the microlens grid instead. This direct strategy is flawed because the requisite raster-scan for full sample coverage would require the input light field to be scanned over a large angular range - 0.33 rad (31°) in our prototype. This large angular span will introduce an unacceptable amount of aberration and significantly deteriorates image resolution. In addition, this required angle would require a bulky and expensive scanning mechanism.

The use of the Talbot effect in the FTM method is predicated on its subtle property: the regenerated Talbot grid can laterally translate a significant amount in response to a small angular tilt of the input light field (Fig. 1(b)) (15). Mathematically, the translation of the first Talbot image (n = 1), $\Delta$, versus angular tilt, $\theta$, is given by

$$\Delta = 2d^2 \tan(\theta) / \lambda \tag{1}$$

In the context of our prototype, an angular tilt of 8.2 mrad will result in a 30 μm lateral translation of the first Talbot grid- enough to allow the full raster-scan over the sample. This angular range can be easily accomplished with a cost-effective MEMS mirror (Mirrorcle).

Fig. 2(a) shows the FTM system. Our FTM prototype uses a CMOS sensor (Aptina, MT9P031I12STM) coated with a green or a red filter. The optical densities of the filters at the excitation wavelength are approximately 6 and the filter thicknesses are ~10 μm (10). A laser beam from the single mode fiber is collimated, then reflected by the

MEMS mirror, and further expanded to provide an angularly-scannable uniform illumination to the microlens grid (Fig. 2 (b)).

**Image resolution**

The resolution, FOV and collection efficiency of a conventional microscope objective are tied to each other. For FTM, the resolution depends on the Talbot focal spot size; the FOV is related to the extent of the focal spot grid and the sensor's dimension; the collection efficiency is related to the physical proximity of the fluorescence sites to the sensor pixels. The FTM design decouples the tradeoff between resolution and FOV that limits conventional microscopy.

The scalar wave simulations verified that the full-range scan angular tilt does not deteriorate the quality of Talbot grid (Fig. 3(a)). To evaluate the system resolution, we used the FTM to image a resolution target mask on an aluminum layer deposited on the CMOS sensor patterned by focused ion beam (FEI, Nova 600). The experiment established the resolution at 1.2 μm (Fig. 3 (b, c)).

**Full-field FTM fluorescence imaging**

To demonstrate that the FTM can be used to image fluorescent biological specimens with microscopic resolution, we performed full-field FTM imaging of genetically-encoded green fluorescence protein (GFP) by using HEK cells line expressing GFP in the nucleus (Fig. 4 (a, b)). We also demonstrated the capability of FTM to image fixed fluorescence human cancer cells (SK-BR-3), which have a cancer marker (Her2) in their membranes labeled by quantum dots (QDots® 625) (Fig. 4 (c, d)).

The scan step size was 0.60 μm in both X and Y direction to satisfy the Nyquist sampling theorem. The dwell time for each step depended on the fluorescent intensity from the sample. With approximately 1 µW on each focal spot, the pixel dwell time for the HEK cell sample and the SK-BR-3 sample was set at 16 ms and 12 ms, respectively for sufficient signal collection, and the total scan time was 50 s and 38 s, respectively. The data readout and storage time for the process, which was mainly limited by the sensor readout speed, was about 4-5 min. The full area of imaging was 3.9 × 3.5 mm$^2$ on the CMOS sensor. The full field image was reconstructed from a stack of 2500 raw images acquired with 128×114 Talbot focal spots, and the reconstructed image contained 6400×5700 pixels. This FOV is more than 100 times larger than that of a 20X objective.

When we applied a uniform illumination and the fluorescence was directly collected on chip, any structure that is finer than the PSF was indistinguishable (Fig. 4 (e)). Using the FTM method, the images were comparable with microscope images taken with a 20X objective (Fig. 4 (f, g)). The cell membrane and sub-nuclear bodies were discernible in the FTM images (Fig. 4 (b, d)).

Discussion

While the current FTM resolution is adequate for a wide range of biological applications, we believe that the resolution can be further improved. While employing a microlens grid that is capable of tighter focusing may appear to be a reasonable solution, it is actually an approach with diminishing returns. The Talbot effect is a paraxial optical approximation: the Talbot image is unable to faithfully regenerate the focal grid when the focal spot size is comparable to the optical wavelength (13). One promising way by which we can achieve resolution improvement would be to design the microlens so that the large angle (high spatial angular frequency) field projections can correctly "phase in" at the Talbot length to generate more tightly focused spots, with the possible compromise of increased excitation background induced by the low frequency components that do not "phase in".

In summary, we have demonstrated a low-cost, compact, chip-scale fluorescence imaging platform, the FTM, to provide a large FOV microscopy solution. This imaging method is enabled by the self-imaging Talbot effect and the magnified lateral shift property associated with angular tuning of the Talbot projection. The ability to easily expand its FOV would be a significant competitive advantage of FTM. In particular, the image acquisition time of conventional scanning systems scales linearly with the imaging area because of the microscope objective's limited FOV. In comparison, the image acquisition time for the FTM would remain the same regardless of the area covered. Due to its compactness, FTM has potential for integration with other on-chip devices. We envision its diverse applications in point-of-care diagnostics, large-scale screens, and long-term automated imaging.

Materials and Methods:

**Cell culture and staining.** The human breast cancer cell line SK-BR-3 was purchased from American Type Culture Collection (ATCC). Cells were cultured at 37 °C, 5% $CO_2$ in McCoy's 5A medium (ATCC) with 10% fetal bovine serum (FBS, Invitrogen). Before experiment, cells were fixed with 4% formaldehyde for 5 min, and incubated sequentially with mouse anti-HER2 (c-erbB-2) antibody (Invitrogen) for 1 h and 5 nM Qdot® 625 goat F(ab')2 anti-mouse IgG conjugate (Invitrogen) for 1 h, then washed with Dulbecco's phosphate-buffered saline (DPBS, Invitrogen).

**Cell sample mounting process.** Before experiment, cells were mounted on a red filter-coated sensor with a 5 mm-diameter coverslip on top. Mineral oil was then filled at the edge of the coverslip to avoid evaporation.

Acknowledgement:

We thank Danielle V Bower (Biological Imaging Center, Caltech) for the help in sample preparation. P.W.S. is an Investigator with the Howard Hughes Medical Institute, which supported this work.

Figure Legends

Fig. 1. The FTM operating principle. (a) The schematic diagram of the FTM system: the microlens grid creates the original focal grid at $Z_f$. The sample on a filter-coated CMOS sensor is located at the Talbot length $Z_T$ away from the original focal grid. (b) As the plane wave is tilted by a small angle $\theta$, the Talbot focal grid laterally shifts by $\Delta$. (c) The scan and reconstruction method for the full-field image.

Fig. 2. (a) The FTM prototype. (b) The diagrams of the Talbot illuminator setup. L1, fiber collimator f=13 mm; L2, f=10 mm; L3, f=50 mm; M, 45° mirrors; MLG, microlens grid; PBS: polarizing beam splitter; ¼WP, quarter-wave plate.

Fig. 3. (a) The scalar wave simulations. (a1, a2) The first Talbot focal grid with incident angle of 0 rad and 4.1 mrad respectively. (a3, a4) The intensity profile of the spot specified by the square in (a1) and (a2), respectively. (b) Resolution target with the structure from 3.0 µm to 1.0 µm: a FIB image of the resolution target (b1); the FTM image (b2). (c) The cross-section line trace of the horizontal (c1) and vertical (c2) marks of 1.2 µm.

Fig. 4. Full field FTM images. (a) HEK cells expressing GFP in nuclei. (b) A magnified view of (a) corresponding to the FOV of a 20X microscope objective. (c-d) SK-BR-3 cells with Her2 marker in the membrane labeled by Qdot®625. (e-f) Images of the same sample by on-chip direct detection using uniform illumination (e), FTM (f), and a conventional microscope with a 20X objective (g).

Fig. 1

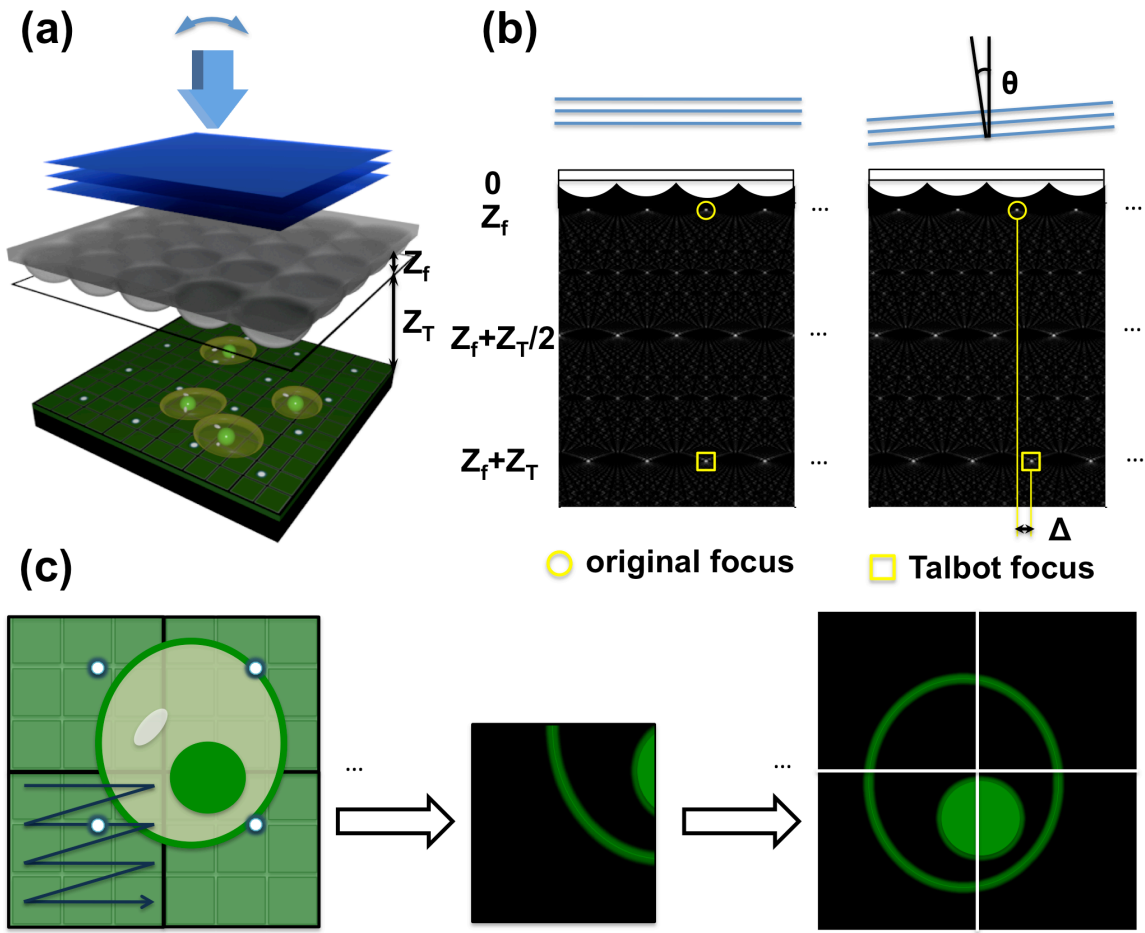

Fig. 2

(a) Talbot Illuminator, Filter-coated CMOS

(b) MEMS Mirror, ¼ WP, PBS, M, L2, L1, M, L3, MLG (M), M, M

Fig. 3

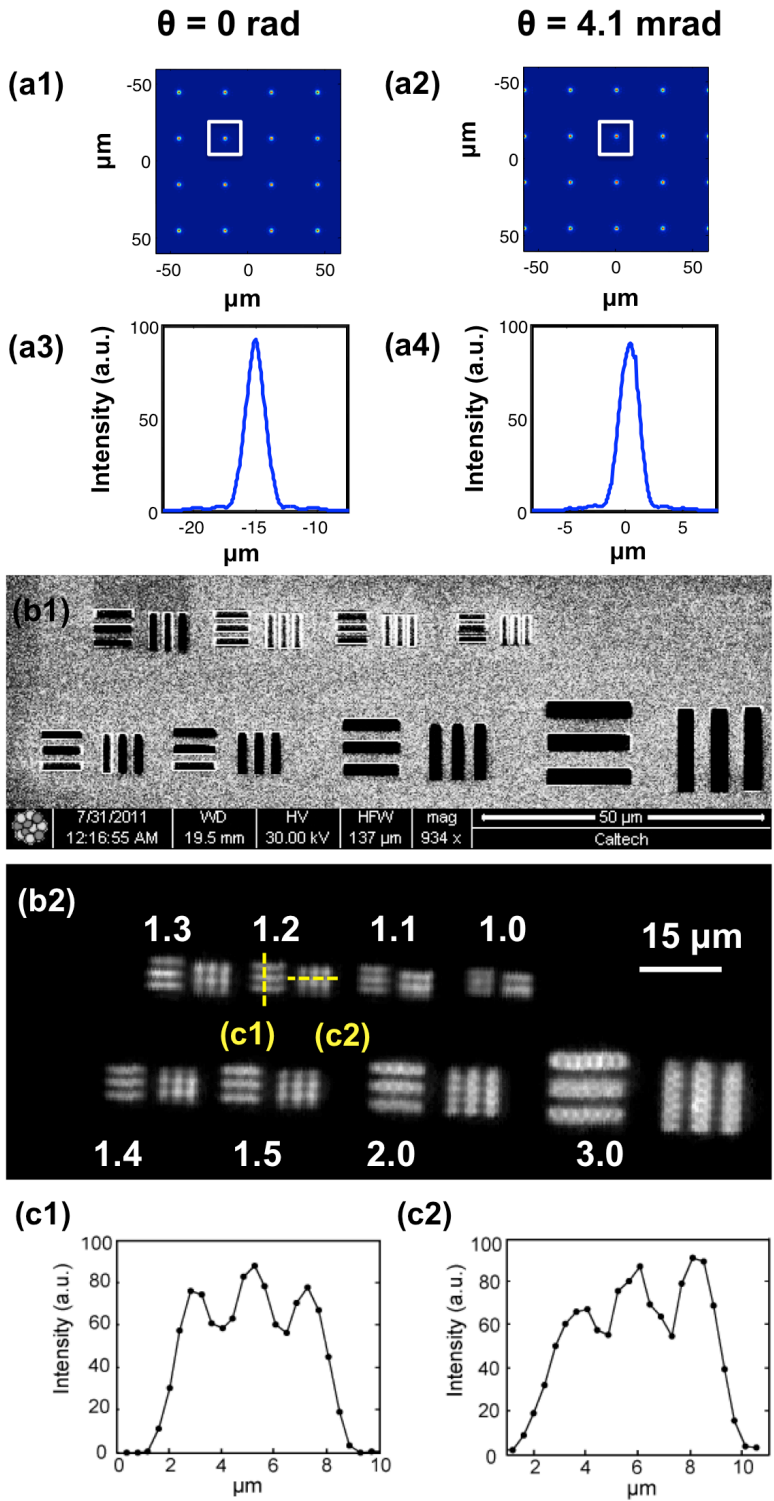

Fig. 4

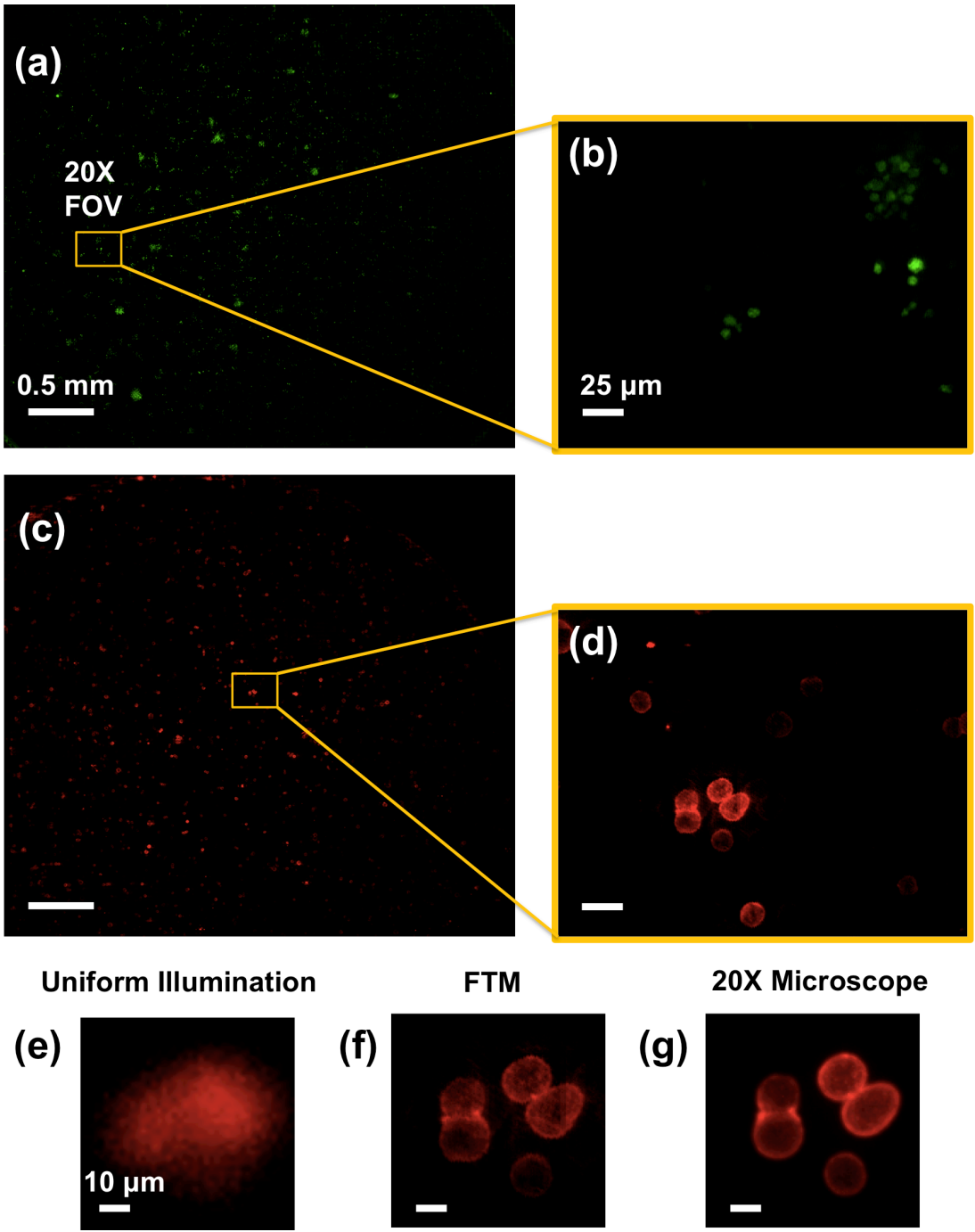